\documentclass[prl,aps,twocolumn,superscriptaddress,nofootinbib]{revtex4-1}
\usepackage{epsf}
\usepackage{epsfig}
\usepackage{graphics}
\usepackage{hyperref}
\begin{document}

\title{Radio frequency spectroscopy of the 
attractive Hubbard model in a trap}

\author{Sanjoy Datta}
\email{sanjoy.datta@grenoble.cnrs.fr}
\affiliation{Harish-Chandra  Research Institute,
 Chhatnag Road, Jhusi, Allahabad 211019, India}
\affiliation{Universit\'{e} Grenoble I and CNRS, 
Laboratoire de Physique et Mod\'{e}lisation 
des Milieux Condens\'{e}s, UMR 5493, B.P. 166, 38042 Grenoble, France}
\altaffiliation[Also at] { Institut N\'{e}el, Universit\'{e} Grenoble I 
and CNRS, B.P. 166, 38042 Grenoble, France} 
\author{Viveka Nand Singh} 
\affiliation{Harish-Chandra  Research Institute,
 Chhatnag Road, Jhusi, Allahabad 211019, India}
\author{Pinaki Majumdar}
\email{pinaki@hri.res.in}
\affiliation{Harish-Chandra  Research Institute,
 Chhatnag Road, Jhusi, Allahabad 211019, India}


\begin{abstract}
Attractive interaction between fermions can lead to pairing and 
superfluidity in an optical lattice. In contrast to the `continuum', 
on a lattice the trap induced density variation can generate a non 
monotonic profile of the pairing amplitude, and completely modify 
the spectral signatures of any possible pseudogap phase.
Using a tool that fully captures the inhomogeneity and strong thermal 
fluctuations, we demonstrate how the crucial radio frequency signatures 
of pairing are `inverted' in a trapped attractive fermion lattice 
compared to the traditional continuum case. These features would be 
central in interpreting any spectroscopic hint of fermion pairing and 
superfluidity.
\end{abstract}

\date{\today}
\maketitle

Optical lattices allow
controllable cold atom realisation 
\cite{jaksch1,essl-rev,bloch-rev,jaksch-hubb} 
of interacting quantum lattice models.
The achievements include the 
observation of a Fermi surface \cite{kohl} and Mott insulating
phase \cite{schneider,jordens} for repulsive fermions,
and the evidence of
superfluidity (SF) \cite{chin}
and anomalous expansion 
\cite{hackermuller} in the attractive case. 
While the canonical antiferromagnetic 
state \cite{gorelik,chiesa} of repulsive fermions
and superfluidity in the attractive 
Hubbard model \cite{hofs} (AHM) remain inaccessible, 
the observation of {\it precursors} to these 
states would already be a major advance.

Even if a pairing induced gapped, or
pseudogap (PG), phase is thermally accessible, the 
spectroscopic signatures would be hard to interpret. 
The well developed theory of  pairing  in the `flat' AHM 
\cite{scalettar,moreo,paiva} provides 
no obvious guidance on the angle resolved spectrum 
of the trapped lattice. The complication has a simple
origin.
Trapping potentials lead to a monotonic
increase in density, as one moves from the edge to the 
center of the trap, but the {\it pairing amplitude }
variation becomes non monotonic once the central density
crosses unity. The non monotonicity affects the spatial
character of excitations, and generates a spectroscopic
response differing drastically from the 
famed `backbending'
that one observes in the flat lattice 
or the trapped continuum~gas~{\cite{rf-refs,pg-obs}}.

\begin{figure}[b]
\vspace{-.2cm}
\centerline{
\includegraphics[width=4.5cm,height=4.5cm,angle=0]{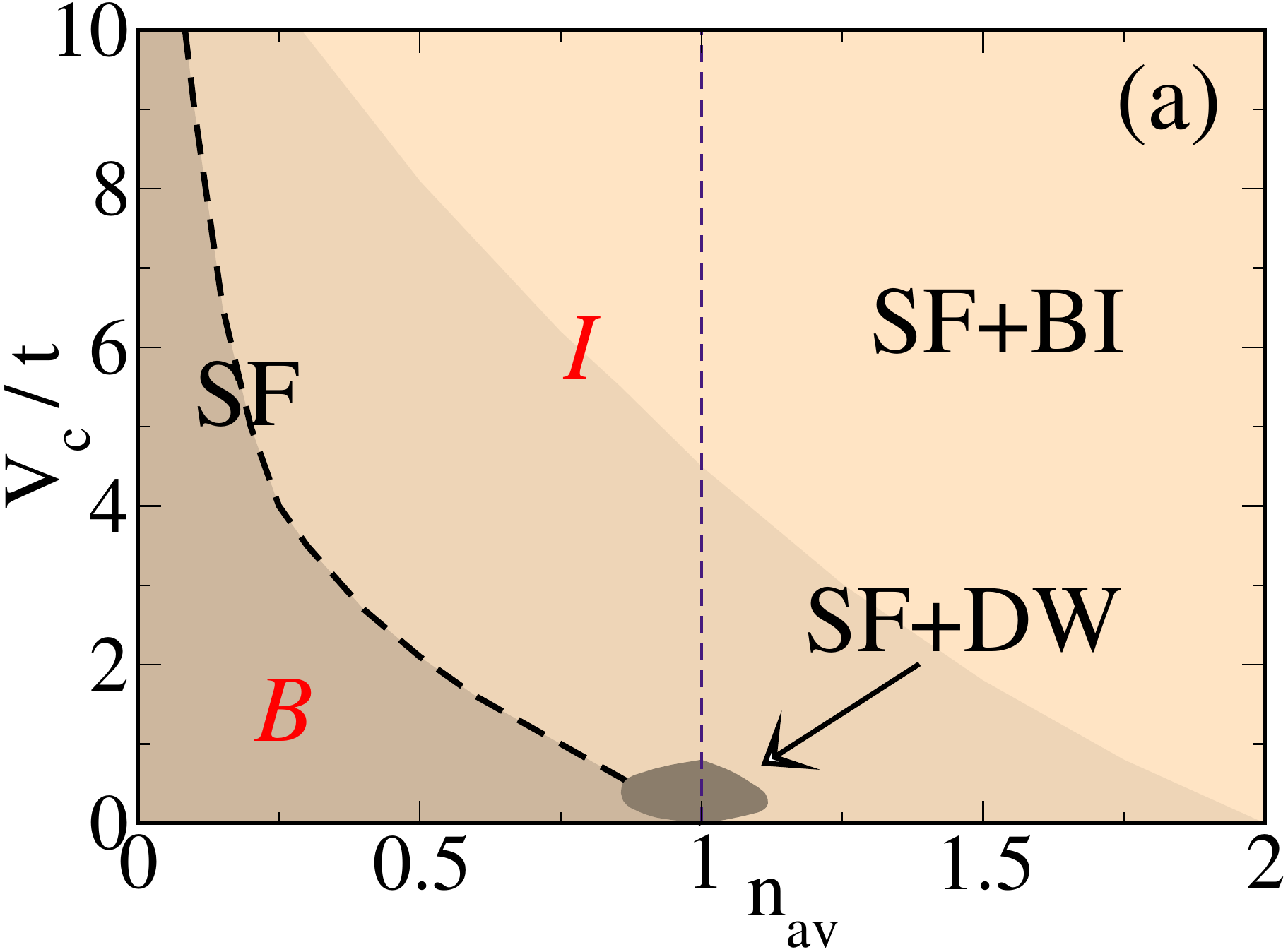}
\hspace{-.1cm}
\includegraphics[width=4.5cm,height=4.5cm,angle=0]{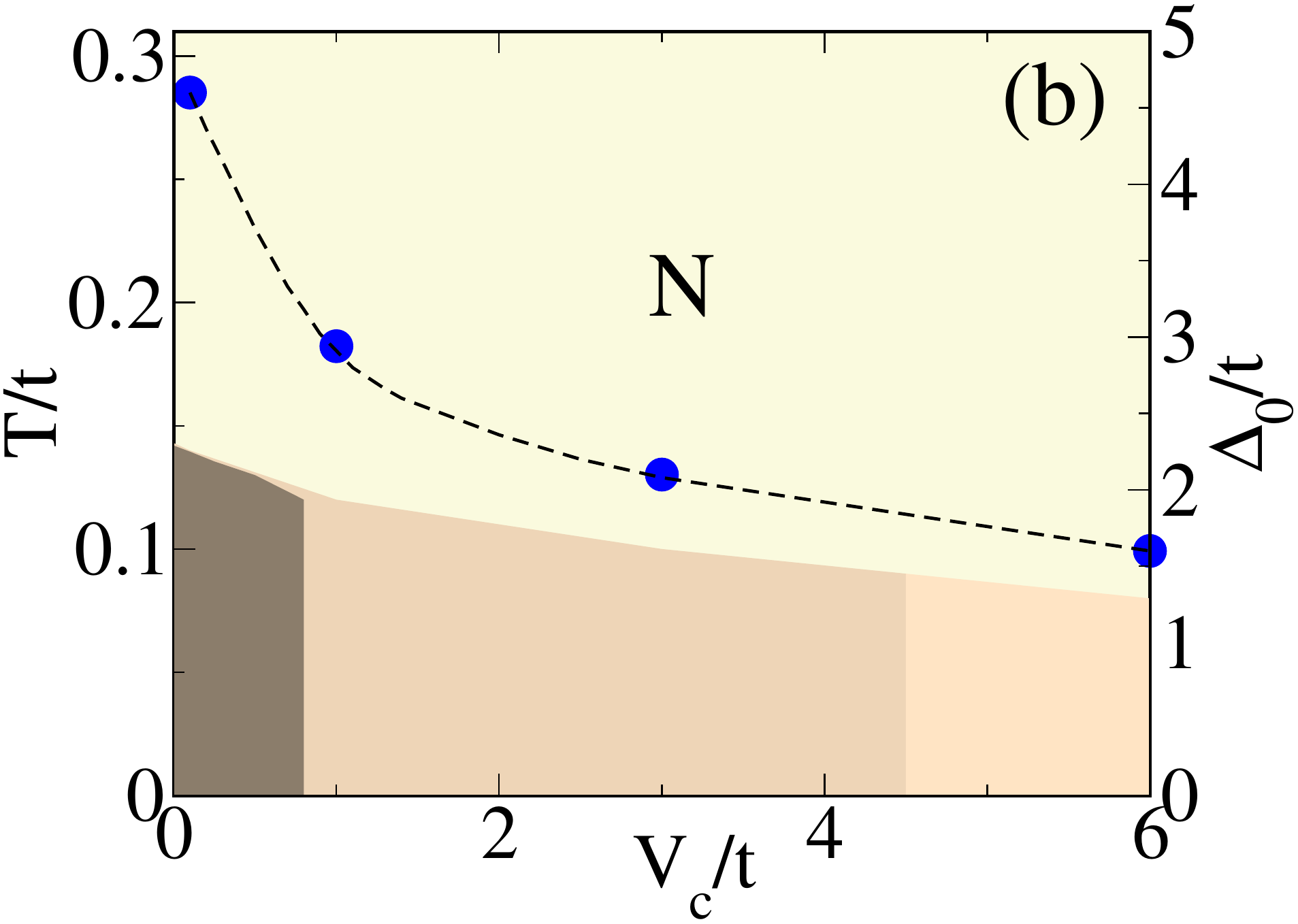}
}
\caption{Colour online: (a)~Ground state of the 2D AHM
at $U/t=6$ for varying average density $n_{av}$ and trapping 
strength $V_c/t$. The tiny region near $n_{av}=1$ corresponds to
strong density wave (DW) 
correlation coexisting with
superfluidity. The band insulator
(BI) refers to the presence of a $n_i=2$ core. 
The $B$ and $I$ regions, separated by the dotted line,
correspond to `backbended' and `inverted' RFS.
(b)~Finite temperature phase diagram at $U/t=6$ and $n_{av}=1$.
Beyond the small window at weak $V_c$ the system has only
SF order at low temperature, with an increasing BI core
for $V_c/t \gtrsim 4$. $N$ refers to the normal state. 
The $T_c$ (on left axis) 
falls monotonically with $V_c$ as does $\Delta_0$ (right axis), 
the $T=0$ gap in the spectrum.
The $T_c$ at $V_c=0$ would vanish in the infinite volume limit,
the results here are for a 
$24 \times 24$ system. 
}
\end{figure}

We completely solve this problem, using a Monte Carlo (MC) 
 method that handles both the inhomogeneity and thermal
fluctuation on large lattices.  We predict the following:
(i)~Increasing confinement leads to rapid decrease in the overall
spectral gap, pushing weight to low
frequency, and quick suppression of the coherence 
feature at the gap edge.
(ii)~Radio frequency spectroscopy (RFS), 
the cold atom analog of 
angle resolved photoemission spectroscopy (ARPES),
 shows `backbending', the traditional signature of
a pairing gap, 
only for weak trapping and low temperature, with 
the momentum dependent gap smallest near ${\bf k} \sim 
{\bf k}_F \sim \{\pi/2,\pi/2\}$. 
For stronger confinement, however, this inverts to
`forward bending' with the gap largest near 
${\bf k} \sim {\bf k}_F $, {\it 
despite the presence of strong pairing}.
(iii)~This `inversion' is {\it generic}, 
and arises when the 
density at the trap center exceeds 1. It survives beyond $T_c$,
but vanishes for $T \gg T_c$.

We provide an analysis in terms of the 
quasiparticle states in the 
trap, and demonstrate an approximate 
``local density'' approach that captures
most of the MC based features and can
yield reliable RF spectra on very large,
experimentally relevant, lattices.

{\it Model and method:}
We study the two dimensional (2D) 
attractive Hubbard model 
in the presence of a harmonic potential:
$ 
H = H_0  - \vert U \vert \sum_{i} n_{i \uparrow} n_{i \downarrow}
$, where $H_0 =
- t\sum_{\langle ij \rangle \sigma} c_{i \sigma}^{\dagger} c_{j \sigma} 
+ \sum_{i \sigma} (V_i - \mu) n_{i \sigma} 
$.
The first term denotes the nearest neighbour tunneling amplitude
of atoms on the optical lattice, the confining potential
has form $V_i = V_0(x_i^2 + y_i^2)$, $\mu$ is the chemical potential,
and $U > 0$ is the strength of attractive on-site interaction.
$x_i$ and $y_i$ are measured in units of lattice spacing $a_0$.
On a $L \times L$ lattice, the
corner value $V_c = V_{\{L/2,L/2\}} = V_0*2*(L/2)^2$.
We use $L=24$.

The spatial variation in mean value, and the
thermal fluctuation about the mean pairing 
amplitude are crucial
in describing the physics of this system. Unbiased calculations
in the homogeneous limit employ determinantal quantum Monte Carlo
(DQMC) \cite{paiva,scalettar,moreo}
to access finite temperature properties. While there are a few
recent calculations using large system size
\cite{chiesa,rost,assmann}, they are focused on thermodynamic
properties and have not touched upon 
the spectral functions of the AHM.

We use a strategy used earlier on moderately sized systems
\cite{dubi,dag}, augmented by a cluster
Monte Carlo technique \cite{tca} that
readily allows access to system size $\sim 30  \times 30$.
We first derive an effective Hamiltonian by decoupling the 
interaction term simultaneously  \cite{hs-chakr}
in the pairing  and density channels via 
a Hubbard-Stratonovich (HS) transformation. 
The exact transformation puts a constraint on the
coupling constants in these two channels
\cite{footnote1}. We choose both couplings to be unity, 
and neglect the time dependence of the auxiliary fields,
to reproduce 
Hartree-Fock-Bogoliubov-de Gennes (HFBdG) theory at $T=0$. 
Our model is:
$ H_{eff}  =  H_0  + H_{coup} + H_{stiff}$, where 
$H_{coup} =
\sum_{i} ( \Delta_{i} c_{i \uparrow}^{ \dagger} c_{i \downarrow}^{ \dagger } +
\Delta_{i}^{ \star } c_{i \downarrow } c_{ i \uparrow }) 
- \sum_{i} \phi_{i} n_{i}$, and
~~~$H_{stiff}=
{1 \over U} 
\sum_{i} (\vert \Delta_{i} \vert^{2} + \phi_{i}^{2})$.
 $\Delta_i = \vert \Delta_i \vert e^{i \theta_i}$ is a complex scalar 
and $\phi_i$ is a real scalar field. 
The inclusion of $\phi_i$ is essential to 
capture the Hartree shift in the inhomogeneous system.
The $T=0$ state
corresponds to solving 
${\delta {\cal E} }/{\delta \Delta_i}=0$ and
${\delta {\cal E} }/{\delta \phi_i}=0$, where
${\cal E}$ is the energy in the $\{\Delta, \phi\}$ 
background, and 
reproduces mean field theory \cite{ghosal}.
Finite temperature 
configurations $\{  \Delta_i, \phi_i\}$ 
follow the distribution $P\{ \Delta_i, \phi_i\}
\propto Tr_{c,c^{\dagger}}e^{-\beta H_{eff}}$
and may fluctuate significantly from the mean 
field state. 

We use the Metropolis 
algorithm to update the 
$\vert \Delta \vert, \theta$ and $\phi$ variables.
This involves solution of the HFBdG 
equation \cite{ghosal,bdg} for each attempted update,
to compute the fermion trace. 
For determining the acceptance of a move 
we solve the 
HFBdG equation on a 
$8 \times 8$ cluster around the update site.
Global properties  like pairing field correlation, 
density of states, {\it etc}, are computed via 
solution of the HFBdG equation on the full $24 \times 24$
system in equilibrium   $\{  \Delta_i, \phi_i\}$
configurations. 
We have checked (see Supplement) that our $T_c$ matches the DQMC estimate
\cite{paiva} over a wide $U/t$ window.

The parameter space of the trap problem involves
$U/t$, $V_c/t$, average density $n_{av}$, and 
temperature $T/t$. 
To keep the effort manageable we set
$U/t=6$, where the $T_c$ in the flat system is
maximum.
We have explored the variation from 
weak to strong confinement over a wide density window
but will show detailed results mainly at $n_{av}=1$.
 
For $V_0=0$ the model is known
\cite{scalettar,moreo}  to have
a SF ground state for $ 0 < n < 2$, except 
at $n=1$ where there is coexistence of SF and DW 
correlations. For $n \neq 1$ the  SF has 
Bardeen-Cooper-Schrieffer (BCS) character at $U/t \ll 1$
and a Bose-Einstein condensed (BEC) form at $U/t \gg1$.  
What is the effect of confinement?

Fig.1.(a) shows the ground state for varying 
$n_{av} = N_f/L^2$, where $N_f$ is the number of fermions, and
corner potential $V_c$.
 At finite $V$ there is
a small window near $n_{av}=1$ where DW correlations
survive, upto $V_c/t \sim 0.8$ \cite{dw-pap}. 
Beyond this window the system has
only SF order. However, the spatial extent of the SF shrinks
with increasing $V_c$ or $n_{av}$ since the central part
of the trap becomes doubly occupied ($n_i=2$) suppressing
$\Delta_i$. 

Fig.1.(b) shows the $V_c-T$ phase diagram at $n_{av}=1$.
There is a narrow SF+DW window at small $V_c$, beyond
which there is only SF order, with the $T_c$ (left axis) 
decreasing quickly with increasing confinement.
The $T=0$ 
spectral gap $\Delta_0$ (right axis) 
falls even more sharply, dropping from $\sim 4.6t$
at $V_c=0$ to $\sim 1.5t$ at~$V_c=6t$. 
\begin{figure}[t]
\centerline{
\includegraphics[width=2.8cm,height=3.5cm,angle=0]{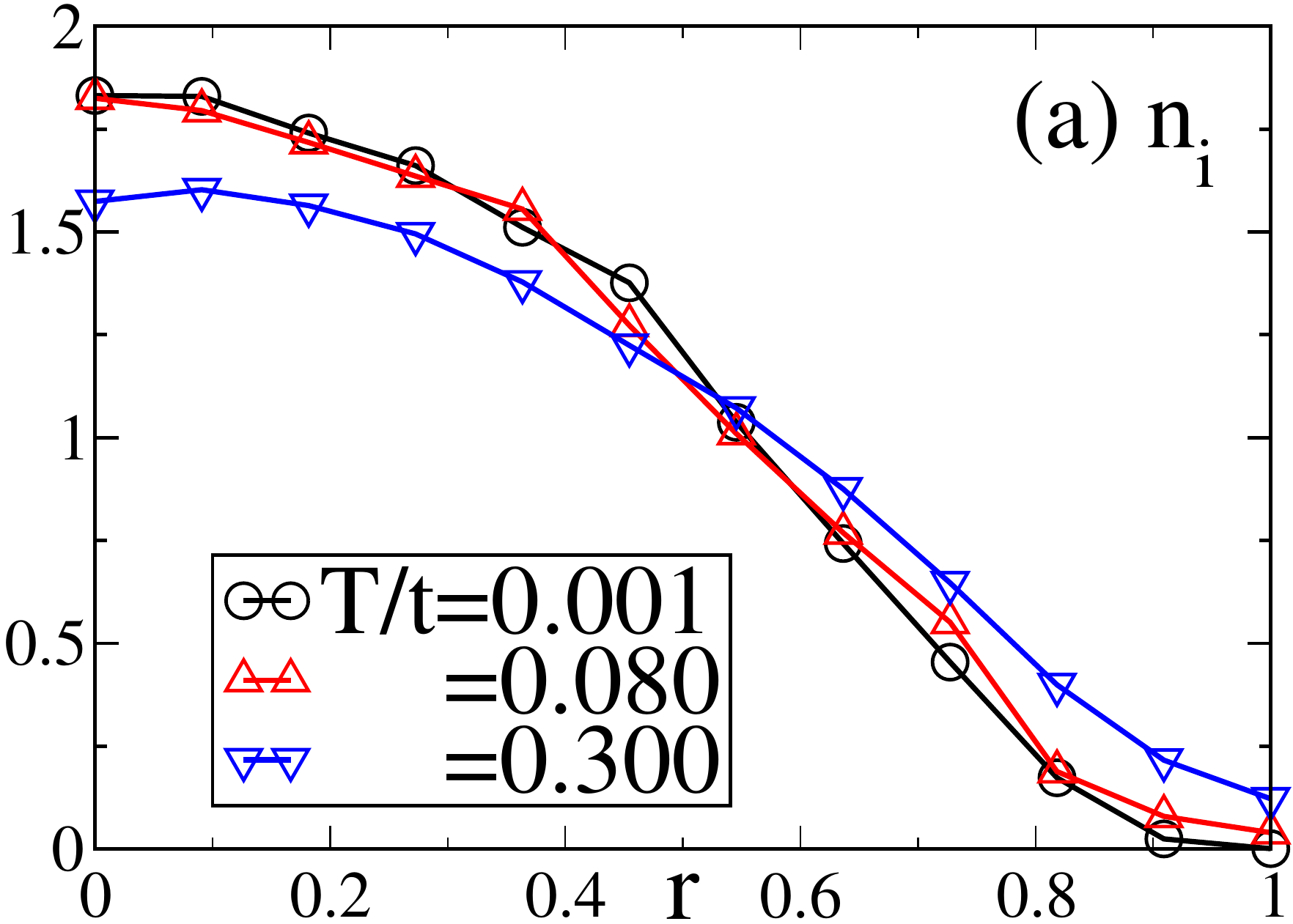}
\includegraphics[width=2.8cm,height=3.45cm,angle=0]{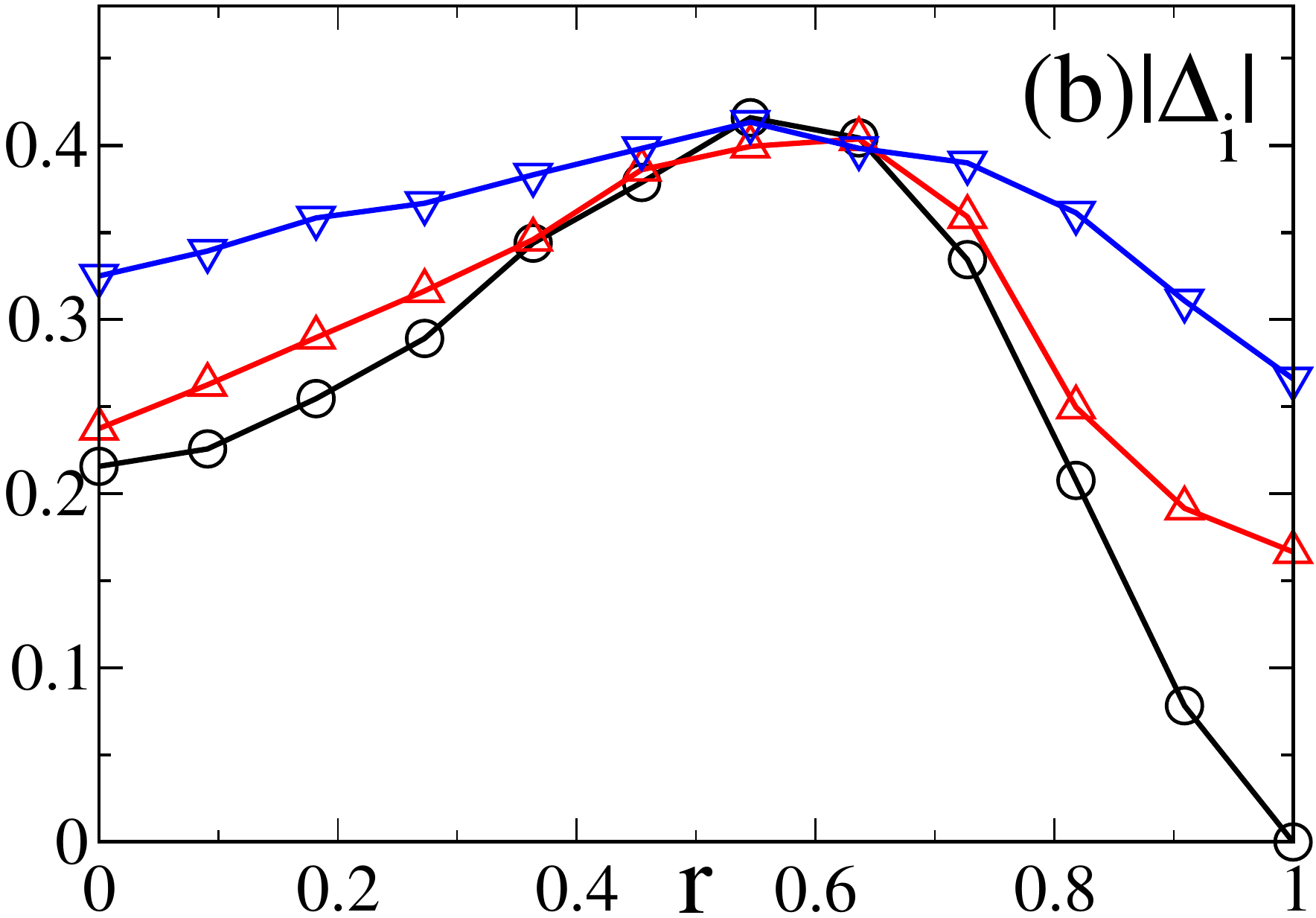}
\includegraphics[width=2.8cm,height=3.45cm,angle=0]{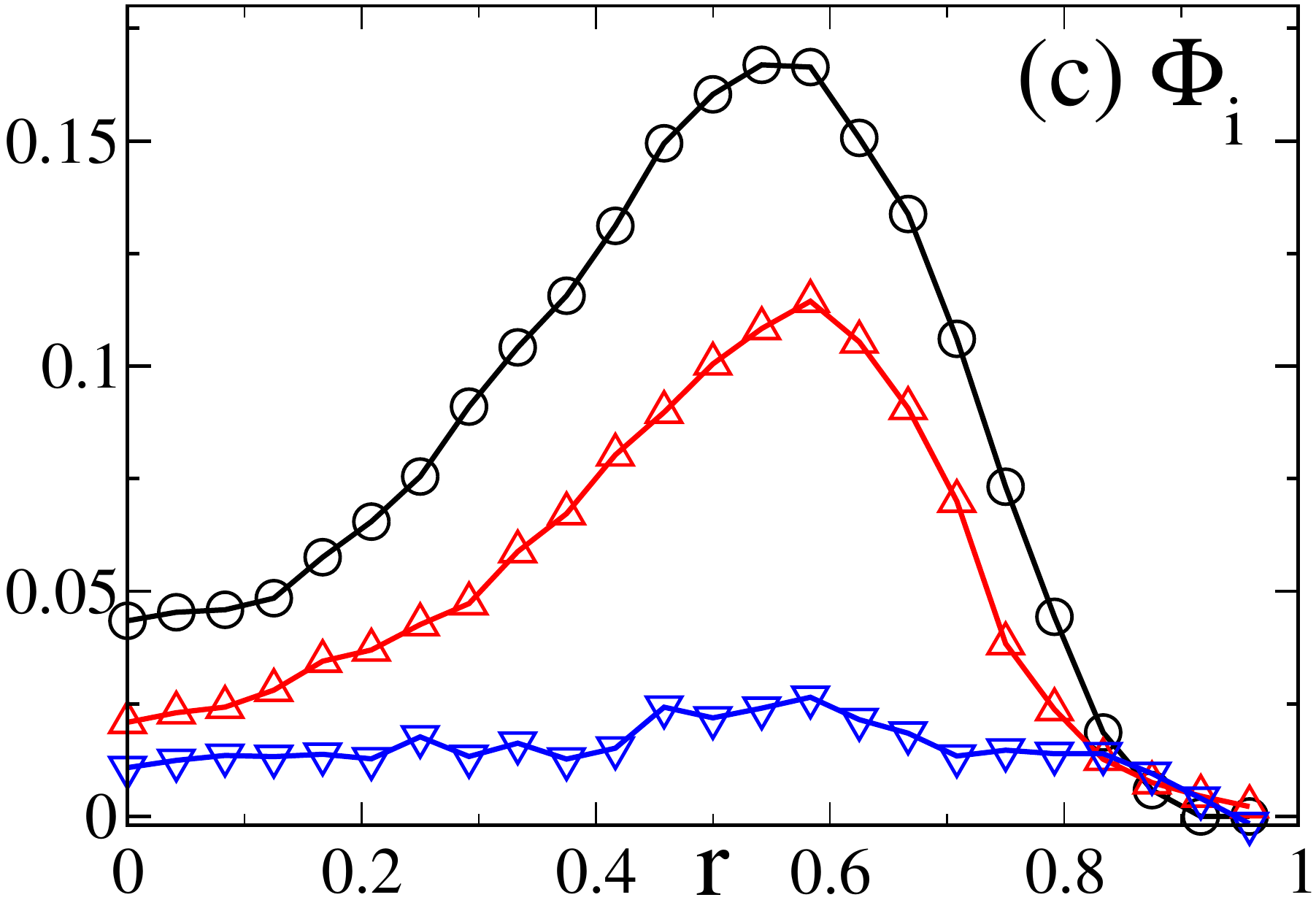}}
\caption{Colour online:
Spatial variation and temperature dependence at
$U=6t$, $V_c=3t$.
(a) density $\langle \langle n_i \rangle \rangle$,
(b) pairing field magnitude $ \langle
\vert \Delta_i \vert \rangle $,
(c) nearest neighbour pairing field correlation.
All patterns are thermally averaged.
}
\end{figure}

Fig.2 shows the radial variation of the thermal average of 
$n_i = \sum_{\sigma} c^{\dagger}_{i\sigma} c_{i\sigma}$
 (left), $\vert \Delta_i \vert$ (center), and 
$\Phi_i = \vert \Delta_i \vert \vert
\Delta_{i + \delta} \vert cos(\theta_i - \theta_{i + \delta}) $
(right).  
The coordinate $i$ is $r = \sqrt{x^2 + y^2}/(L/2)$, varying along
the diagonal. $\Phi_i$ tracks nearest neighbour
correlation in that direction. 
We have set $V_c=3t$ and $n_{av} \sim 1$ and $T=0,~0.08t,~0.3t$.
The full 2D spatial maps are shown in the Supplement.

Fig.2(a) shows the expected monotonic fall in 
$\langle \langle n_r \rangle \rangle$ at all $T$. The
cloud at $T=0.3t$ is slightly broader than at $T=0$.
The pairing field amplitude in 2(b) 
is more interesting.
It is non-monotonic at all $T$, a peculiarity of the
lattice where it grows with $n$ till
$n=1$ and falls beyond.  The $T=0$ result for 
$\langle \vert \Delta_r \vert \rangle$ is what 
is expected from mean field HFBdG theory, with a 
clear peak in the region where $n_r \sim 1$. 
At $T=0.08t$ the amplitude profile looks similar to $T=0$,
but with a large growth in the corner where it was zero
at $T=0$! The trend amplifies at $T=0.3t$ where $\langle
\langle \vert \Delta_r \vert \rangle \rangle $ is
much less inhomogeneous than at $T=0$.
This is due to the low amplitude stiffness in regions
with low $\vert \Delta_i \vert$ at $T=0$. We provide a
connection to the flat system physics in the Supplement.

Fig.2(c) is meant to highlight the suppression of
phase correlation with temperature. At $T=0$ the phases
are locked, so $\Phi_i = \vert \Delta_i \vert \vert
\Delta_{i + \delta} \vert \approx \vert \Delta_i \vert^2$.
At $T=0.08t \sim 0.7T_c$ while the amplitudes are not 
very different from $T=0$ the phase correlation is weakened.
By $T=0.3t$ while amplitudes have grown, NN phase correlations
have weakened to about $20\%$ of the $T=0$ value. Long 
range phase correlation is of course lost at $T_c$.
The  spatial characteristics are not directly accessible so
we move to the spectral signatures that RF spectroscopy can probe.

\begin{figure}[t]
\centering{
\includegraphics[width=2.8cm,height=3.2cm,angle=0]{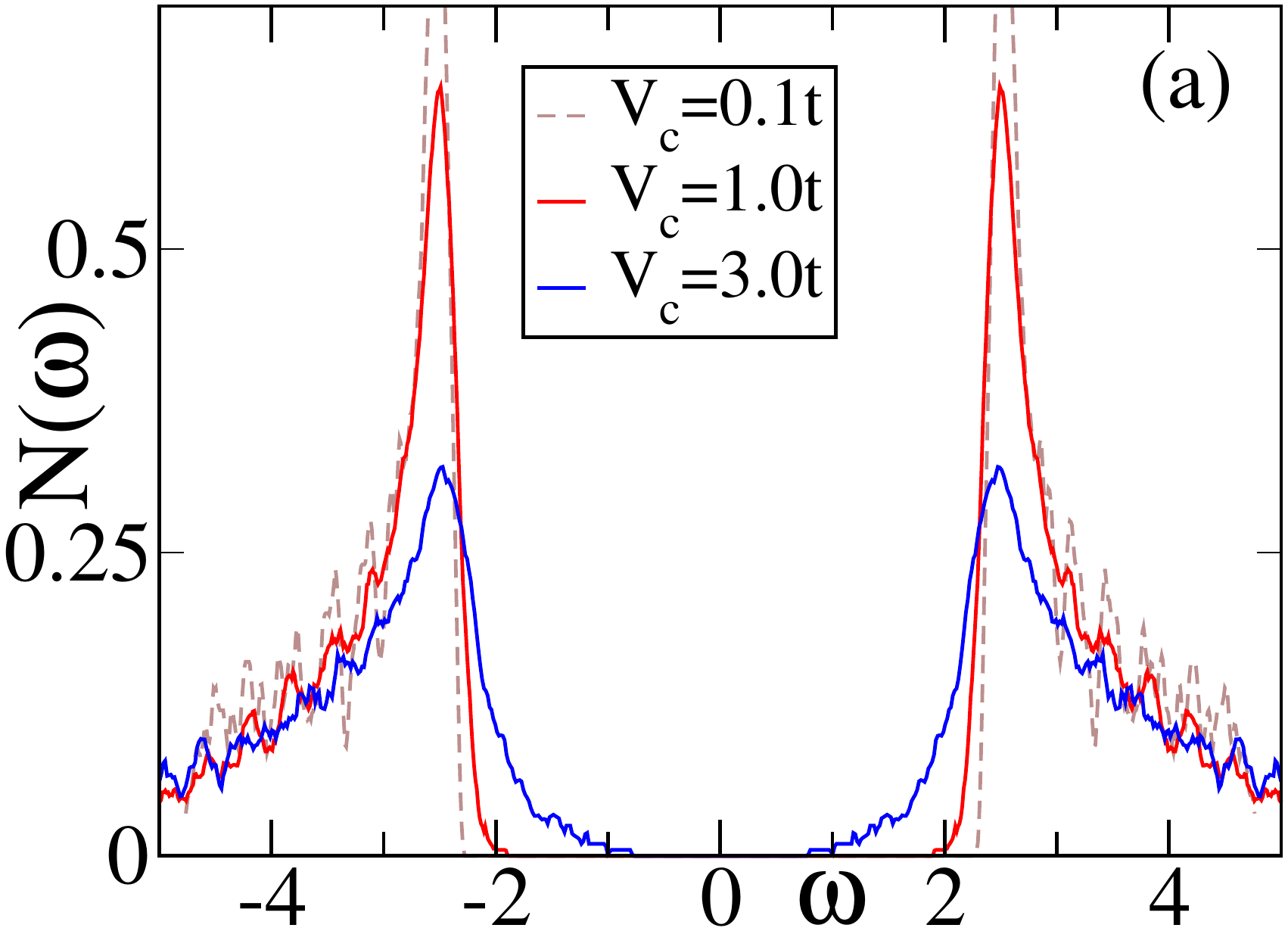}
\includegraphics[width=2.8cm,height=3.2cm,angle=0]{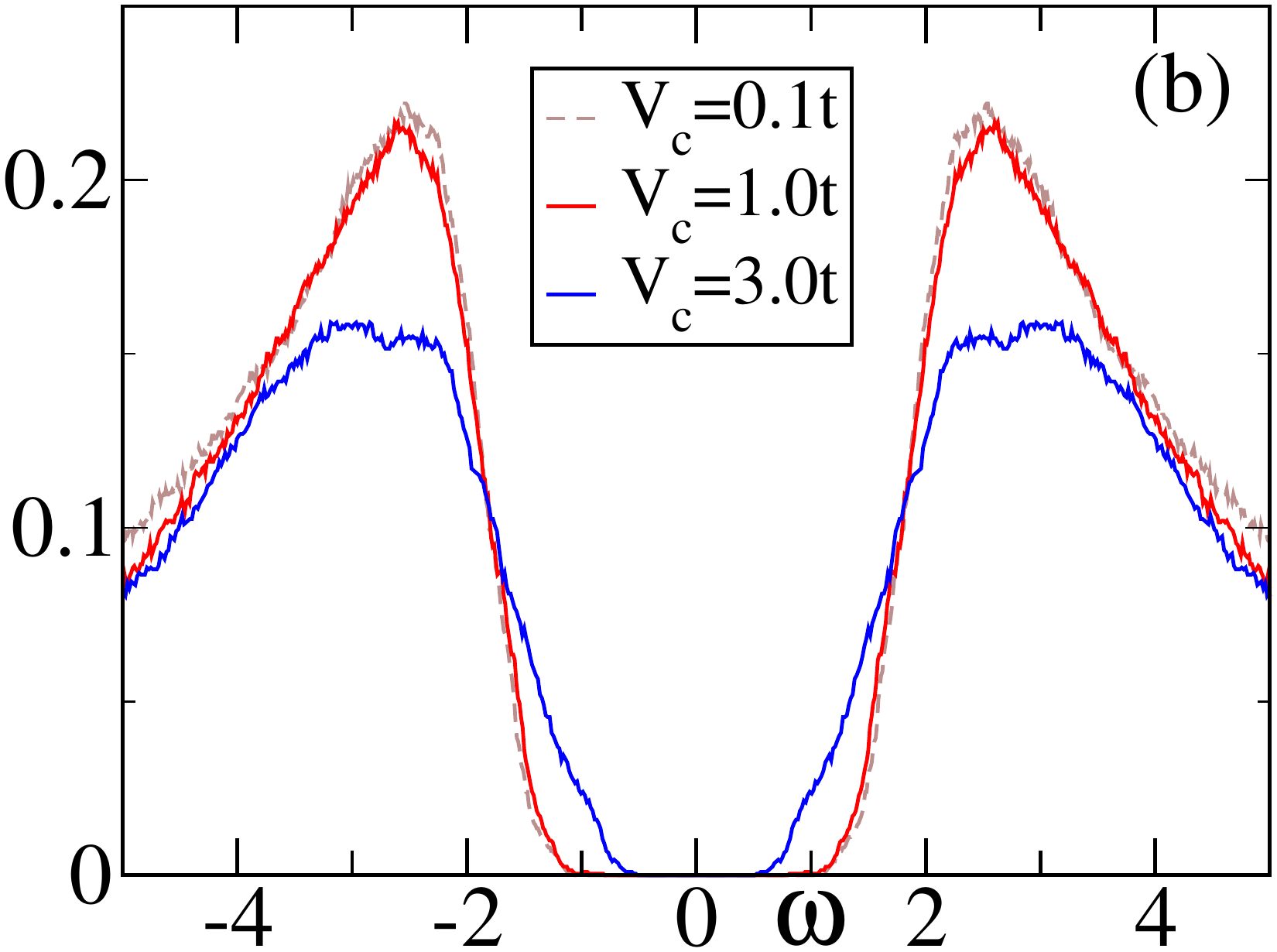}
\includegraphics[width=2.8cm,height=3.2cm,angle=0]{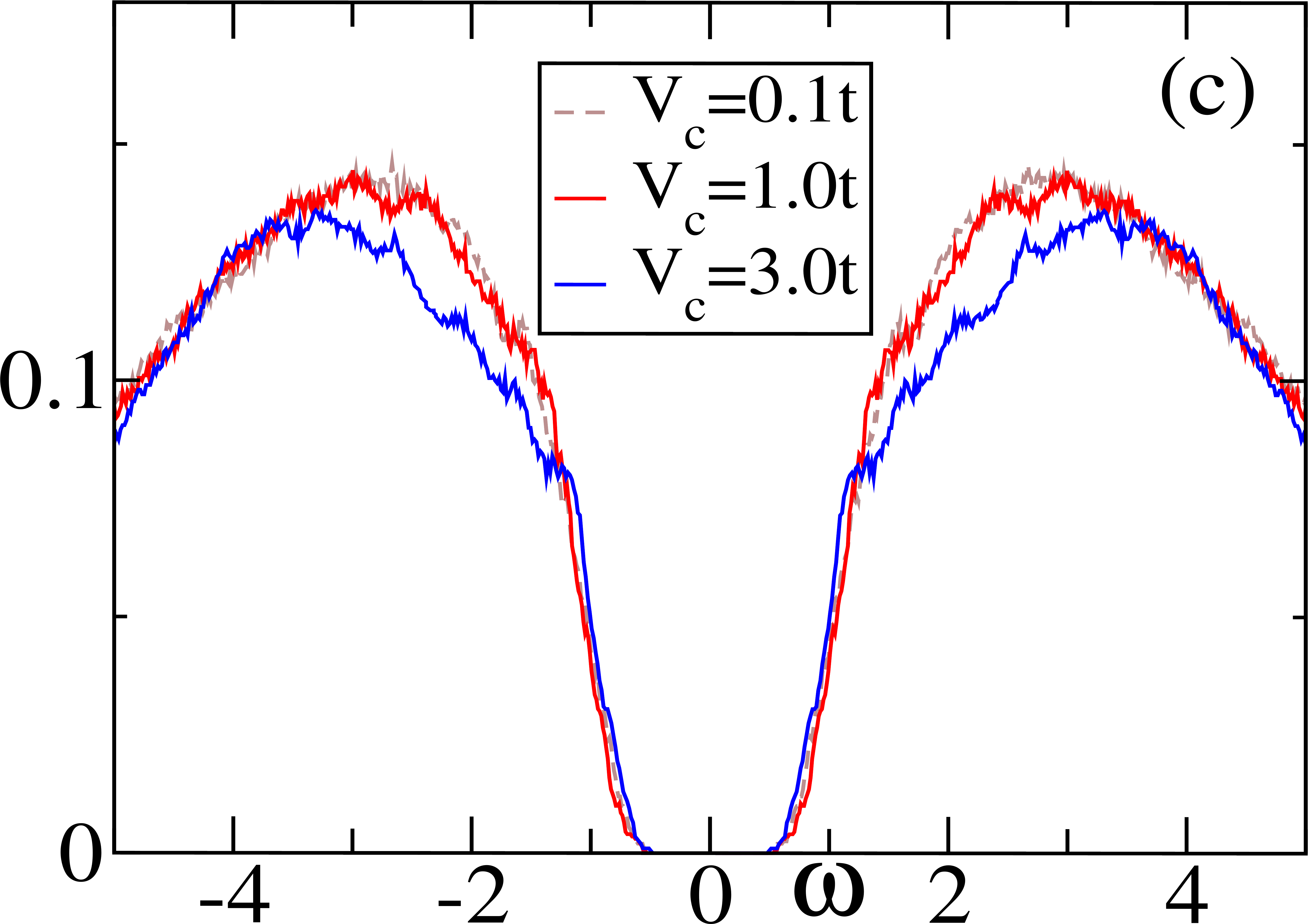}}
\caption{Colour online:
DOS for increasing degree of confinement at three temperatures 
(a)~$T=0$, (b)~$T=0.08t$, and (c)~$T=0.3t$.}
\end{figure}

Fig.3 
shows the single particle density
of states (DOS).
Fig.3.(a) shows $V_c$ dependence at $T=0$.
There are two 
primary effects of trapping: (i)~the effective gap 
reduces with increasing $V_c$ due 
to appearance of low frequency spectral weight,
and (ii)~the `coherence peak' and sharp gap edge 
are blurred.
The decrease in the gap
arises from the smaller pairing amplitude in regions
which have density $n_i \rightarrow 0$ or $n_i \rightarrow 2$.
We have explicitly checked this 
from the local density of states
(LDOS).  In fact at $n \sim 1.9$ the pairing gap in the
flat system is $0.8t$, not
very different from the threshold that we observe.
The $n_i \sim 1$ region 
contributes to spectral weight at $\vert \omega \vert
\gtrsim 2.5t$, consistent with results from  the flat system.
In a flat system the threshold, $\omega_{gap}$, and the
coherence peak location, $\omega_{coh}$, coincide.

\begin{figure}[t]
\centerline{
\includegraphics[width=9.0cm,height=9.0cm,angle=0]{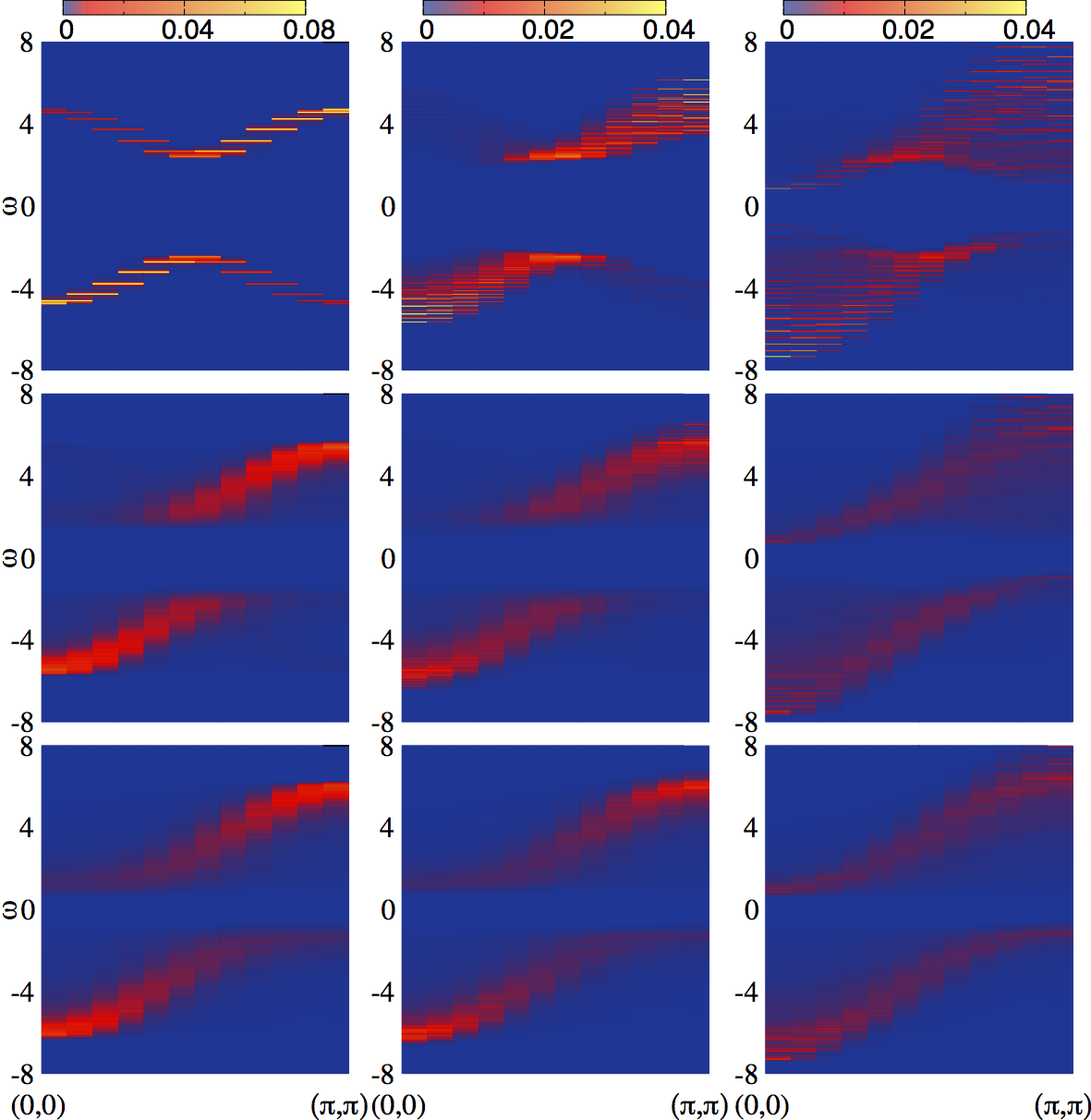}}
\caption{Colour online:The spectral function $A({\bf k}, \omega)$ 
for a `diagonal scan' from ${\bf k} = \{0,0\}
\rightarrow \{\pi,\pi\}$. Along the row, $V_c$ varies from $0.1t,~t,~3t$
(left to right). Down the column $T$ varies from $0,~0.08t,~0.3t$.
The $T_c$ of the unconfined system is $\sim 0.14t$, at $V_c = 3t$ it
is $ \sim 0.1t$.}
\end{figure}

At $T \sim 0.08t$, Fig.3.(b), the DOS for $V_c=0.1t$ and
$V_c=t$ look very similar, with a reduction of $\omega_{gap}$
from the $T=0$ value and suppression of
the coherence peak. The $V_c=3t$ case also shows
reduction of $\omega_{gap}$ with respect to $T=0$, but
continues to be distinct compared to the weaker $V_c$ 
cases. Since $\langle \vert \Delta_i \vert \rangle$
has not changed significantly with respect to $T=0$
(Fig.2.(b))
these changes are attributable to phase disorder. 

By the time $T=0.3t$, Fig.3.(c), 
the DOS in the three cases are essentially similar,
since the $\langle \vert \Delta_i \vert \rangle$ 
homogenises even in the trap (Fig.2).
The density does continue to be inhomogeneous,
affecting $\phi_i$, but $\vert \Delta_i \vert$ 
is more important for
the low frequency spectrum.

The momentum resolved spectral function, Fig.4,
 is more
dramatically affected by trapping.
The $3 \times 3$ panel shows the
spectrum $A({\bf k}, \omega)$. The formal definition in
terms of HFBdG eigenstates is given in the Supplement. 
In each panel, the $x$-axis corresponds to the
${\bf k}$ scan from $\{ 0,0 \}$ to $\{ \pi,\pi \} $, the
$y$-axis is the frequency $\omega$, and
$A({\bf k}, \omega)$ is colour coded as indicated.
The columns are for $V_c =0.1t,~t,~3t$ (left to right),
the rows are $T= 0,~0.08t,~0.3t$ (top to bottom).
The {\it size dependence} of our results is shown in the
Supplement.

The left column at $V_c=0.1t$ shows the
thermal evolution in an essentially flat system.
(i)~Top panel: ground state.
Here $A({\bf k},\omega)
\approx u_{\bf k}^2 \delta(\omega - E_{\bf k})
+ v_{\bf k}^2 \delta(\omega + E_{\bf k})$, where
$E_{\bf k} = \sqrt{(\epsilon_{\bf k} - \mu)^2 + \Delta^2}$, 
$u_{\bf k}$ and $v_{\bf k}$ are the usual
BCS coherence factors, $\epsilon_{\bf k}$ is the tight binding
dispersion and $\Delta$ is the uniform pairing amplitude.
The two dispersing bands correspond to $\pm E_{\bf k}$
and one observes the expected `backbending' in the lower
curve near ${\bf k} \sim \{\pi/2,\pi/2 \} $ \cite{singer}, 
where, for us, $\epsilon_{\bf k} \approx \mu$. 
(ii)~Middle: at $T=0.08t$ coherent
particle-hole mixing is almost lost.
For 
${\bf k} \sim \{\{0,0\} \rightarrow \{\pi/2,\pi/2\}\}$
the spectrum is mainly `particle-like',
while for 
${\bf k} \sim \{ \{\pi/2,\pi/2\} \rightarrow \{\pi,\pi\}\}$
it is `hole-like'. There is significant mixing only near
${\bf k} \sim \{\pi/2,\pi/2\}$. There is a 
faint surviving trace of the mean
field, $\pm E_{\bf k}$, dispersion, the $+E_{\bf k}$ 
branch for ${\bf k} \sim \{0, 0\}$ and the $-E_{\bf k}$ 
branch for ${\bf k} \sim \{\pi,\pi\}$.
Effectively there are three branches in $A({\bf k},\omega)$
at each ${\bf k}$.
(iii)~Bottom: at $T = 0.3t$ there is no
trace of the mean field $E_{\bf k}$, 
the spectrum is an incoherent combination of
upper and lower band features at~all~${\bf k}$.

For $V_c=t$, middle column, 
the moderate confinement already shows signatures in
$A({\bf k},\omega)$. (i)~Top panel: the  $T=0$
spectral functions are broad since ${\bf k}$ states
overlap with multiple trap eigenstates. 
Low ${\bf k}$ states have large (and broad) weight
in the lower band while ${\bf k} 
\sim \{\pi,\pi\}$ involves
broad weight in the upper band.
The gap between
the upper and lower bands is still smallest at 
${\bf k} \sim \{\pi/2,\pi/2\}$ and the  
backbending feature has not vanished.
(ii)~At $T=0.08t $ and $T=0.3t $ the results are
similar to what we saw for the flat case, with some
extra (trap induced) broadening noted above.

The third column shows results at $V_c=3t$
where the trap center density is $n_i \approx 2$.
The ARPES differs qualitatively from the flat case.
(i)~Top: at $T=0$ the $A({\bf k}, \omega)$ is
very broad since a large number of trap eigenstates overlap
with $\vert {\bf k} \rangle$.
The interband separation now has a {\it maximum} for
${\bf k} \sim \{\pi/2,\pi/2\}$ and is minimum for ${\bf k}
\rightarrow \{0,0\}$ or $\{\pi,\pi\}$.
This is a case of `forward bending' rather than
backbending.
If RF spectroscopy probes the edge of the lower band it
would obtain a concave pattern, rather than the convex 
result that traditionally indicates a pairing gap.
The gap, as is obvious from the full $A({\bf k}, \omega)$ 
is nevertheless present.
(ii)~Middle: at $T=0.08t $ all gaps are smaller compared to
$T=0$ but the unusual ${\bf k}$ dependence persists.
(iii)~Bottom: at $0.3t $ there is only the hint of the
${\bf k}$ dependent gap observed at lower $T$.
How do we relate these results to~spatial~structure?

The overall DOS is $N(\omega) = -(1/\pi) Im \sum_i G_{ii}(\omega)$, 
{\it i.e}, a sum of the local DOS over the system
where
$G_{ii}(\omega)$ is the local projection of the spin averaged
fermion Green's function. If the density $n_i$ were {\it slowly }
varying then as a starting approximation we could
use $G_{ii}^{trap}(\omega) 
\approx G^{flat}(\omega,n=n_i)$.
We have checked that this works reasonably even on 
our $24 \times 24$ system. The overall DOS is
then given by
$N(\omega) \approx \int dn P(n) N_{flat}(\omega,n)$, where
$N_{flat}(\omega,n)$ is the flat system DOS at density $n$ and
the density distribution 
$P(n) = {1\over N} \sum_i \delta(n - n_i)$ can be 
computed from the MC density profile.

This immediately creates a connection between the density
(and auxiliary field) variation in the trap and the features
observed in the DOS. 
The ARPES, however, involves the overlap $\langle {\bf k}
\vert m\rangle$ of a plane wave state with a BdG eigenstate
$\psi_m$. 
If all $\psi_m$ were extended over the
system, and overlap all $\vert {\bf k} \rangle$, the strange
gap modulation with ${\bf k}$ would not arise.

We find that 
the BdG states are {\it radially 
localised to a remarkable degree}, the Supplement shows
typical real space and momentum space patterns.
The lowest energy excitation at $T=0$, at $E_m \sim 0.9t$, is
localised near the corners, where $n_i \rightarrow 0$.
This has fourier modes only near ${\bf k}
=\{0,0\}$. For $E_m \gtrsim 1.3t$ the excitations shift
to the center of the trap, and involve modes near ${\bf k}
\sim \{\pi, \pi\}$. Only for  $E_m \gtrsim 2.5t$, where the
BdG states have large weight on the $n_i \approx 1$
annulus do we see contribution at ${\bf k} \sim
\{\pi/2,\pi/2\}$.

Although our system size is larger than accessible in
typical DQMC studies, it is well below the $\sim 100 \times
100$ lattices used in experiments. This is where the 
local density approximation (LDA) to $P(n)$ becomes
useful.
LDA prescribes that 
$n_i^{trap} \approx n_{flat}(\mu_i)$, where 
$\mu_i = \mu - V_i$ and $n_{flat}(\mu)$ 
can be computed from DQMC or analytic approximations. 
In the Supplement we compare the MC based ARPES data with
results obtained using LDA on the same size. The agreement
is remarkable. We extended this to a huge $\sim 200 \times 200$
system, and all the qualitative features of our original result
survive.

{\it Conclusions:}
We provide the first solution to the angle resolved 
spectral properties of an attractive fermion lattice 
in the presence of confinement, crucial for any
cold atom experiment. Even a moderate trapping 
potential creates a `core' with low pairing amplitude
and generates spectral features that are widely
different from the well studied `continuum' model
and the `flat' Hubbard lattice.
We point out a novel `forward bending' feature that
would be the RF spectroscopy signature
 of a pairing gap, clarify the spatial 
origin of this feature, and illustrate a scheme that 
allows access to 
the spectrum on very large experimentally
realised lattices.

{\it Acknowledgments:}
We acknowledge use of the Beowulf Cluster at HRI.
PM acknowledges support from a 
DAE-SRC Outstanding Research Investigator
Award.


%
%
%
\begin{center}
\large{Supplementary material for\\
{\bf ``Radio frequency spectroscopy of the attractive 
Hubbard model in a trap''}}
\end{center}

\begin{center}
\large{Sanjoy Datta, Viveka Nand Singh and Pinaki Majumdar} 
\end{center}

\vspace{.1cm}

\begin{center}
\textbf{I.~Benchmarking the auxiliary field Monte Carlo}
\end{center}
\begin{figure}[ht]
\centering
\includegraphics[height=5.5cm,width=8cm]{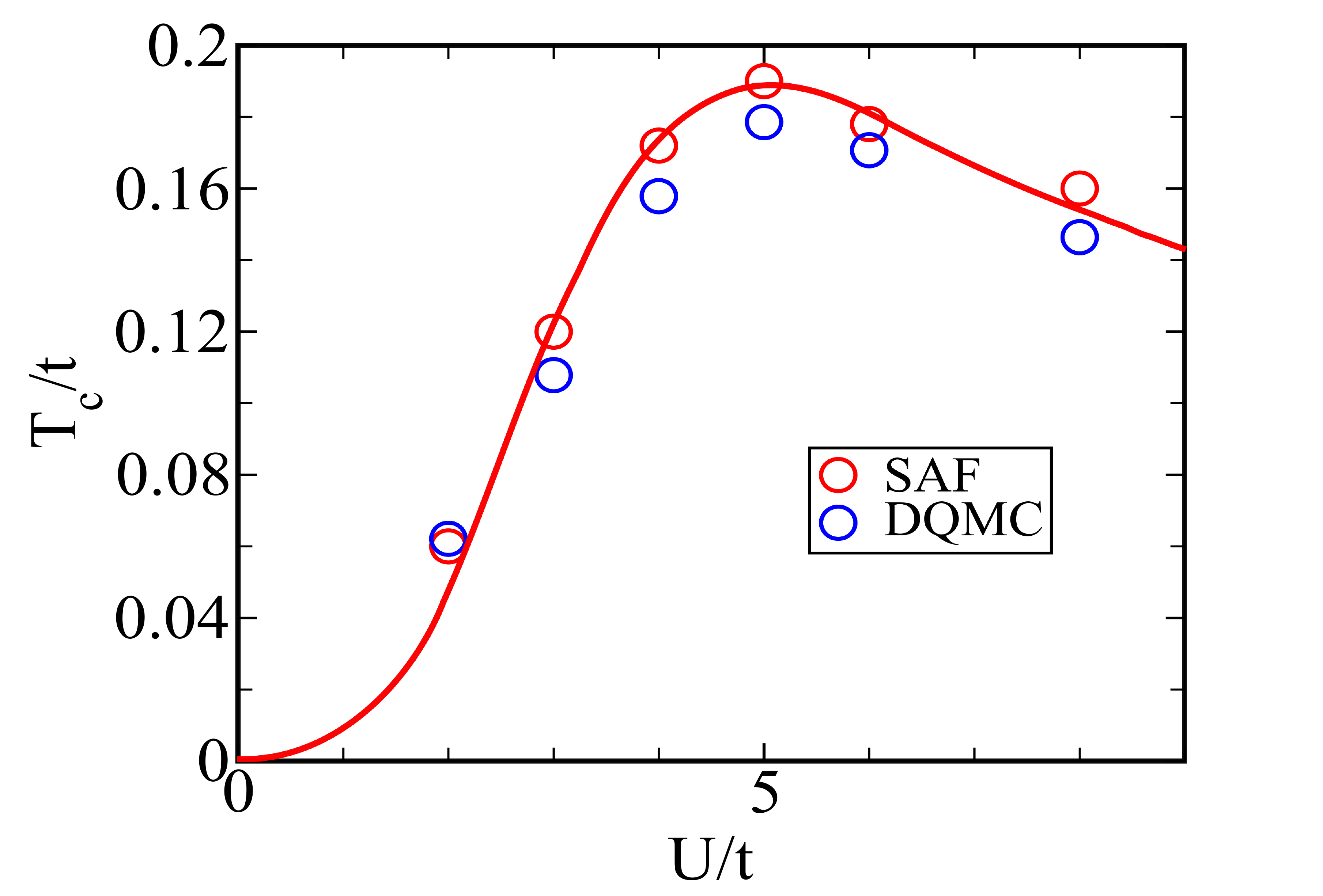}
\caption{Comparison of superfluid $T_c$ obtained within our static
auxiliary field (SAF) scheme, with DQMC (Ref. 15 of main text).
The results are for a
flat system with density $n = 0.7$ and lattice size
 $10 \times 10$.  }
\label{benchmark}
\end{figure}
Large scale determinantal Monte Carlo (DQMC) results are not available for the
trapped problem so we compared the results of our method to DQMC data in the
`flat' problem.
DQMC results for the superfluid transition temperature ($T_c$) are
available at density 
$n=0.7$ on a $10\times10$ lattice for $U/t$ varying from $2-8$.
Fig.\ref{benchmark} compares our results to this benchmark. 
We capture the non monotonic character, the correct peak location,
and our $T_c$ estimate 
is within $10\%$ of the DQMC result at all $U/t$. 
This is far superior to mean field theory which would have generated a $T_c$
growing monotonically with $U/t$, with an {\it order of magnitude
overestimate} already at $U/t=6$.
%
  
\begin{center}
\textbf{II.~Homogenisation of $\langle \vert \Delta_i 
\vert \rangle $ with growing temperature}
\end{center}
\begin{figure}[h]
\centering
\includegraphics[height=6cm,width=8cm]{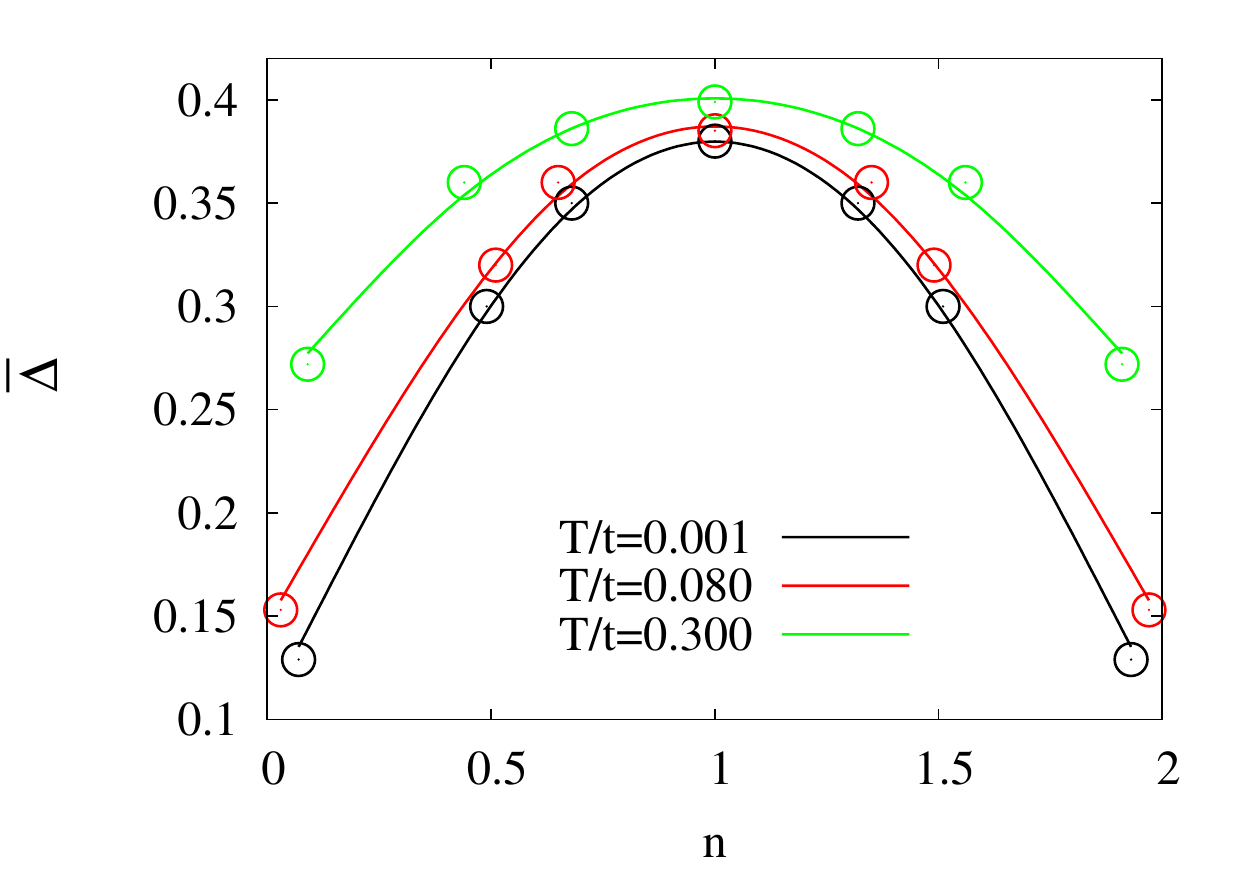}
\caption{The thermal average of $\vert \Delta \vert$ on a `flat' system for
varying density $(n)$ and temperature $(T)$. 
The $T=0.001t$ result corresponds to mean field theory, the finite $T$
results involve fluctuations within the SAF scheme.
}
\label{avgdelta}
\end{figure}
We have seen in Fig.2 of the main text that the mean value
of $ \vert \Delta_i \vert  $ tends to 
become independent of position ${\bf r}_i$ with growing $T$,
even though the density $n_i$ remains inhomogeneous.
We found that this is related to the lower amplitude stiffness
of regions with low $\vert \Delta \vert $ at $T=0$, and has a direct 
correspondence with the behaviour in flat systems. 
We studied the mean value ${\bar \Delta}(n,T) = 
(1/N) \sum_i \langle 
\vert \Delta_i \vert \rangle_{n,T}$ in the
flat system and discovered that although ${\bar \Delta}$
is strongly $n$ dependent at $T=0$, with a $70\%$ variation
as $n$ changes from $1.0$ to $0.1$, at $T=0.3t$ that
variation is only $\sim 30\%$. 
This flat system effect shows up in the trap as a local
amplitude stiffness that depends on the $T=0$ magnitude
of $\vert \Delta \vert$ in that region.

\begin{center}
\textbf{III.~Behaviour of Bogoliubov-de Gennes (BdG) wavefunctions}
\end{center}
%
\begin{figure*}[h]
\centering
\includegraphics[height=1.5in,width=1.6in]{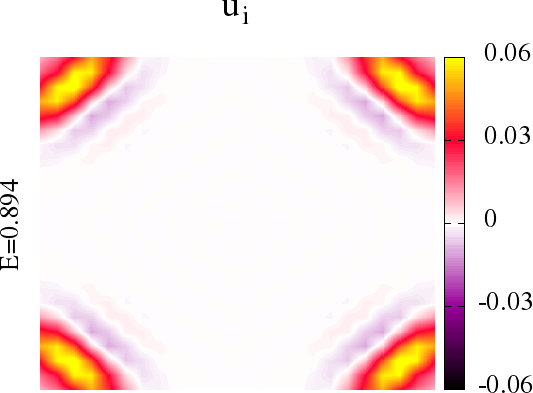}
\includegraphics[height=1.5in,width=1.6in]{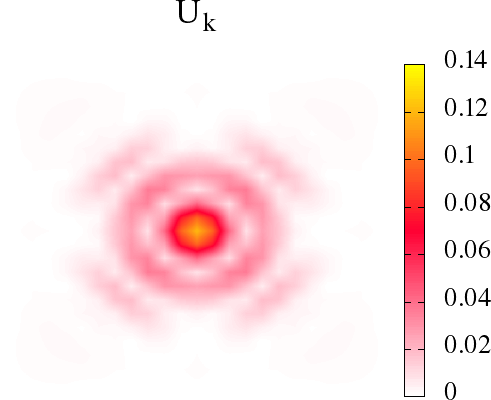}
\includegraphics[height=1.5in,width=1.6in]{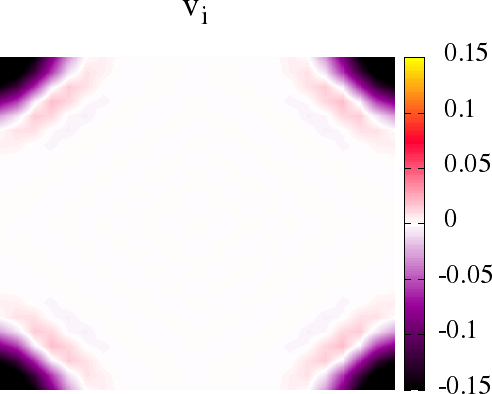}
\includegraphics[height=1.5in,width=1.6in]{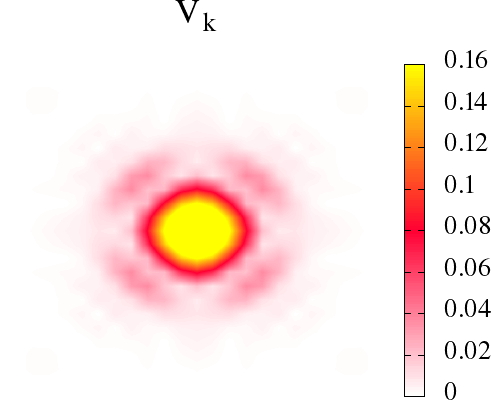}\\
\includegraphics[height=1.5in,width=1.6in]{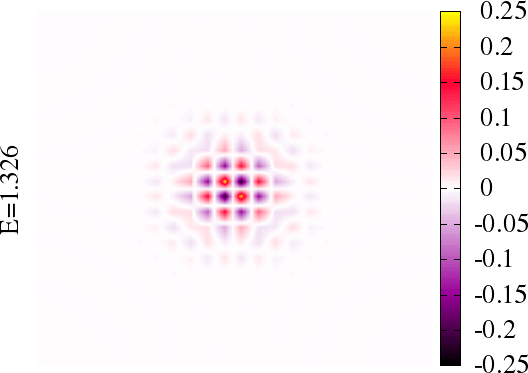}
\includegraphics[height=1.5in,width=1.6in]{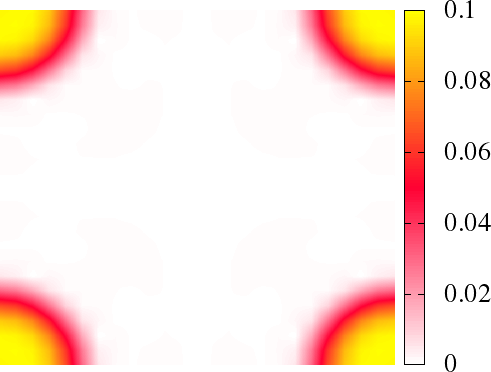}
\includegraphics[height=1.5in,width=1.6in]{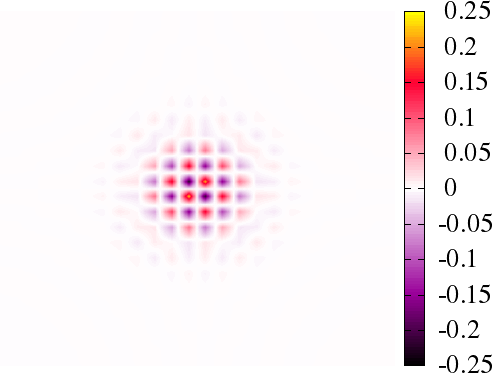}
\includegraphics[height=1.5in,width=1.6in]{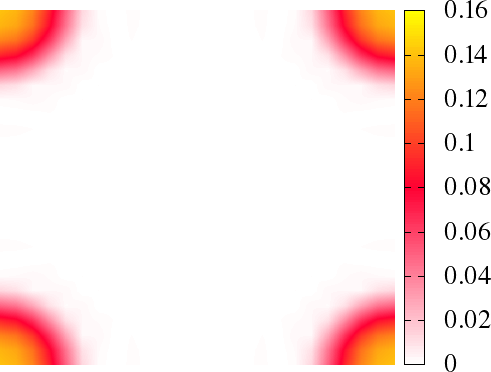}\\
\includegraphics[height=1.5in,width=1.6in]{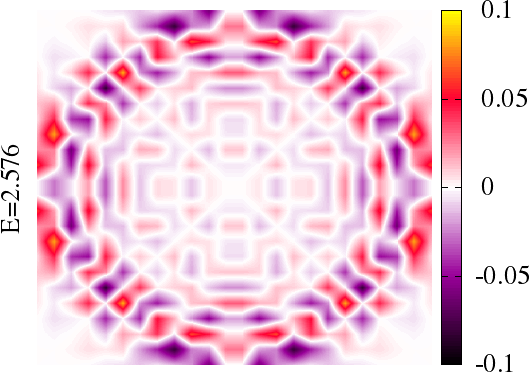}
\includegraphics[height=1.5in,width=1.6in]{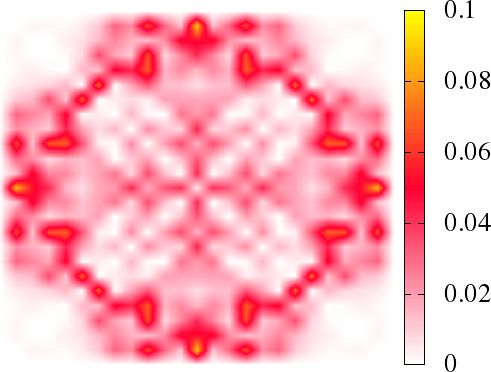}
\includegraphics[height=1.5in,width=1.6in]{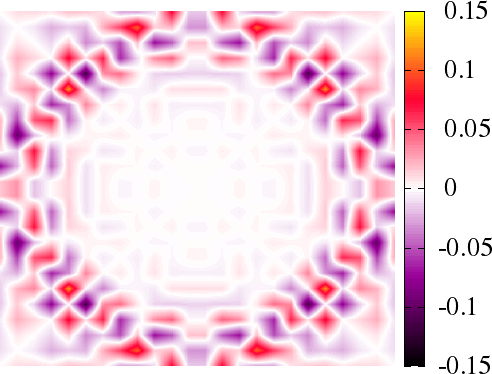}
\includegraphics[height=1.5in,width=1.6in]{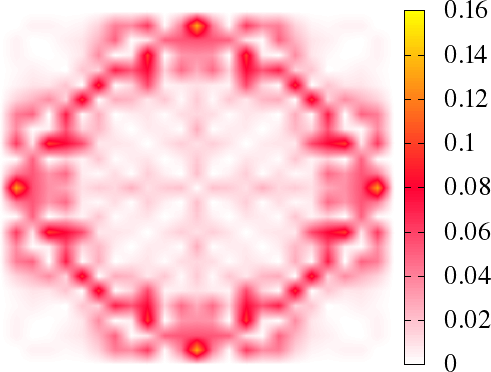}\\
\includegraphics[height=1.5in,width=1.6in]{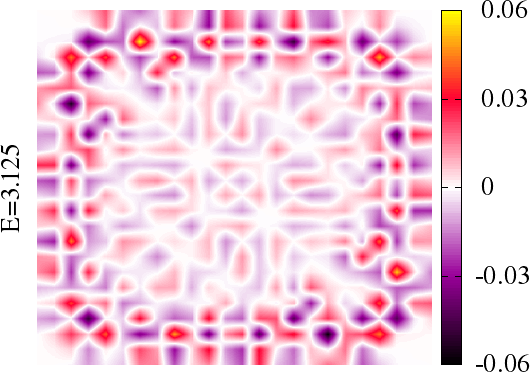}
\includegraphics[height=1.5in,width=1.6in]{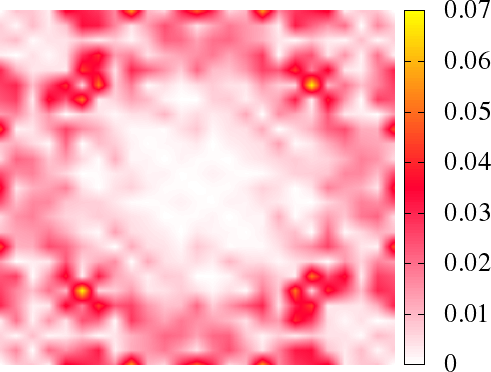}
\includegraphics[height=1.5in,width=1.6in]{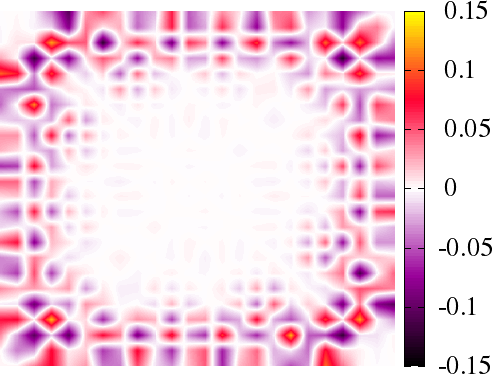}
\includegraphics[height=1.5in,width=1.6in]{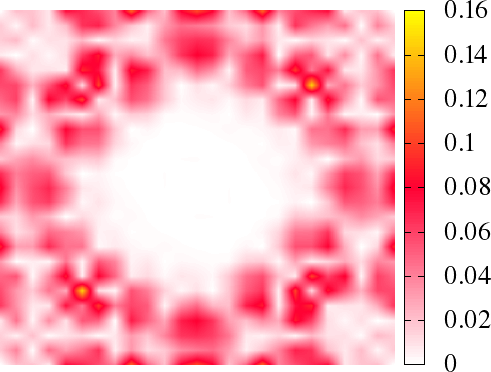}
\caption{BdG eigenfunctions for 4 different excitation energies.
Along the row $u^m_i,~U^m_{\bf k},~v^m_i,~V^m_{\bf k}$. Along the column
different $m$, starting with the lowest $E_m$. 
}
\label{bdg}
\end{figure*}
We  analysed  the BdG wavefunctions in real space and in terms of
their
momentum content, and show a few illustrative examples in 
Fig.\ref{bdg} for
$V_c = 3t $.
BdG wavefunctions in real space have been represented by $u_i,v_i$ and in 
momentum space as $U_k,V_k$. 
The center of the spatial maps is ${\bf r}_i =(0,0)$. For the momentum maps
the center is ${\vec k} =(0,0)$ and the corners are $(\pm \pi,\pm \pi)$.
{\bf A.}~The 1st row of 
Fig.\ref{bdg} corresponds to the lowest energy excitation.  
One can see that (i)~the state has large amplitude in the
low density region at the corners 
and, (ii)~$U_k$ and $V_k$ are
large near $\vec k =0$.
The low gap in $A({\vec k}=0,\omega)$ 
arises due to overlap with this excitation.  
{\bf B.}~Second row, 
$E=1.326t$, higher up in the spectrum.
This state is (i)~mainly localised at 
the center of the trap, {\it i.e}, the highest density region
and  (ii)~is the first state with 
significant $\vec k = (\pi,\pi)$
content.
The $\vert \Delta_i \vert$ here is small, but larger 
than in the corner region.
{\bf C.}~Rows 3 and 4 show states with contribution at
$\vec k=(\pi/2,\pi/2)$. These are spread over the system but have
significant weight in the $n \approx 1$ annulus, where 
$\vert \Delta \vert$ at $T=0$
is largest.
The states are at significantly higher energy than
the states in rows 1 and 2.
%

\begin{center}
\textbf{IV.~Computation of the momentum resolved spectral function.}
\end{center}
The spectral function $A(\vec k,\omega)$ for a given configuration of 
$\Delta_i, \phi_i$ has been calculated via the following expression
\begin{equation}
A(\vec k,\omega) = \sum_{n,E_n \ge 0} \left[ \left|u_n(\vec k)\right|^2
\delta(\omega - E_n) + \left|v_n(\vec k)\right|^2
\delta(\omega + E_n)\right], \nonumber
\end{equation}
where
\begin{eqnarray}
u_n(\vec k)&=&\frac{1}{N^{1/2}} \sum_{i=1}^N e^{i \vec k \cdot \vec r_i} 
u_n(\vec r_i) \nonumber \\  
v_n(\vec k)&=&\frac{1}{N^{1/2}} \sum_{i=1}^N e^{i \vec k \cdot \vec r_i} 
v_n(\vec r_i) \nonumber .
\end{eqnarray} 
To get the final $A(\vec k,\omega)$ it has been averaged over many 
equilibrium configuration of $\Delta_i, \phi_i$.
\begin{center}
\textbf{V.~Comparison of exact and local density approximation
(LDA) based spectral functions} 
\end{center}
\begin{figure*}[h]
\centering
\includegraphics[height=1.5in,width=1.5in]{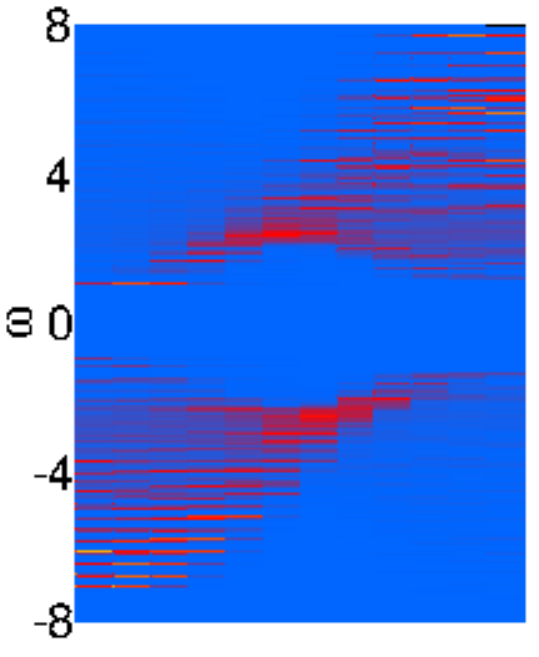}
\includegraphics[height=1.5in,width=1.4in]{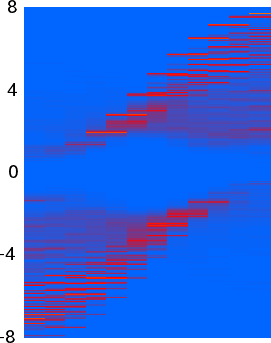}
\includegraphics[height=1.5in,width=1.4in]{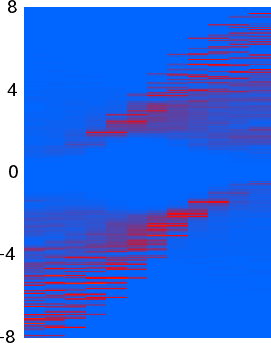}
\includegraphics[height=1.5in,width=1.6in]{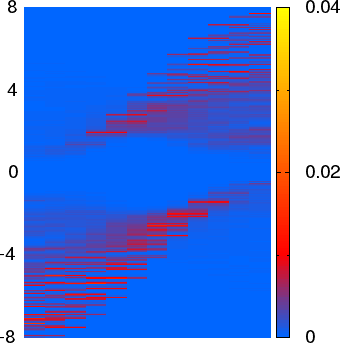}\\
\includegraphics[height=1.6in,width=1.5in]{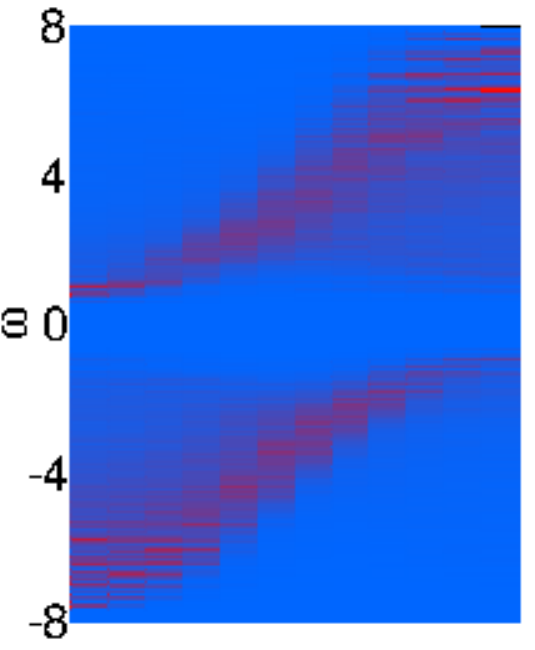}
\includegraphics[height=1.5in,width=1.4in]{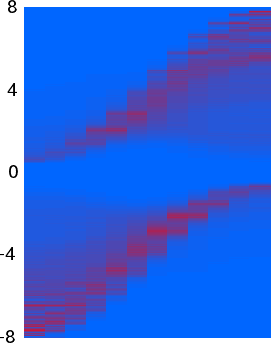}
\includegraphics[height=1.5in,width=1.4in]{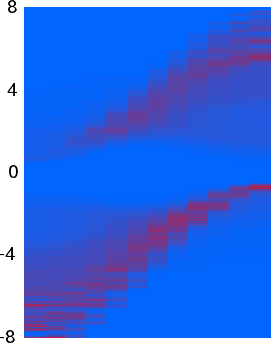}
\includegraphics[height=1.5in,width=1.6in]{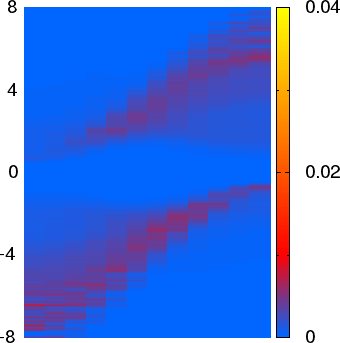}\\
\includegraphics[height=1.6in,width=1.5in]{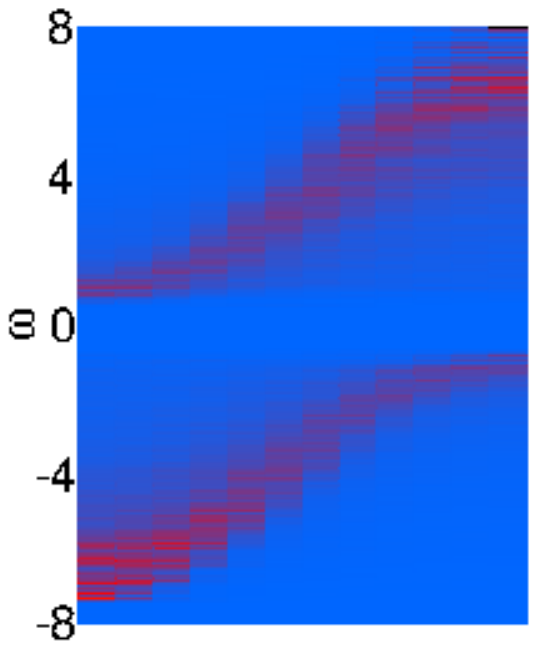}
\includegraphics[height=1.5in,width=1.4in]{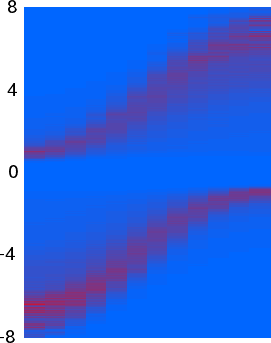}
\includegraphics[height=1.5in,width=1.4in]{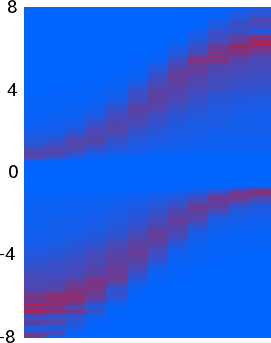}
\includegraphics[height=1.5in,width=1.6in]{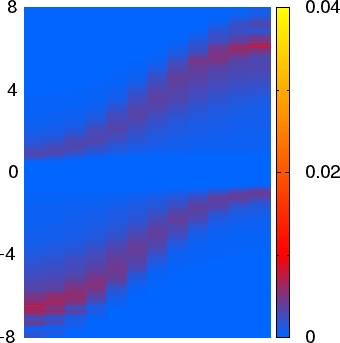}
\caption{Comparison of the actual $A(\vec k,\omega)$ (first column) with that
based on $P(n)$ obtained from the full calculation (second column), and on
$P(n)$ obtained
from LDA scheme for $24\times24$ lattice (third column) and $192 \times 192$
lattice (fourth column).}
\label{lda}
\end{figure*}

Pushing the `local density' approach to the momentum resolved spectral function
we checked the accuracy of this approach in capturing
$A({\vec  k},\omega)$ in the trap.
We computed the `local density'  based spectral function $A^L_{trap}(\vec k,\omega)$
as follows: 
\begin{equation}
A^L_{trap}(\vec k,\omega) = \int P(n)  A_{flat}(n, \vec k,\omega) dn,
\label{eqn_lda}
\end{equation}

This prescription is incomplete without specifying $P(n)$. The first
approximation is to use the $P_{MC}(n)$ that emerges from the MC itself.
This approach, although it does not require BdG based information,
 still requires MC generated data, and is impractical on large sizes,
$\sim 100 \times 100$, that are likely to be used in experiments.
For that $P(n)$ itself needs to be approximated. 

We tested the standard prescription that,
for a slowly varying density
field, one can relate $n_i$ to a {\it local chemical potential}
$\mu_i = \mu - V_i$, where $n_i$ and $\mu_i$ are related by the
same equation of state as in the homogeneous system. That relation
we infer from numerical results on the flat system. 
The $\mu_{LDA}$ itself is fixed by
requiring ${1 \over N} \sum_i n_i(\mu_i) = n_{av}$.
From $n_i$ one can generate the `local density
approximation' result $P_{LDA}(n)$.
This can be computed easily on any size, and we generated it
on $24 \times 24$ and $192 \times 192$ lattices.

Fig.\ref{lda} compares the `exact' spectral function at
$V_c=3t$  with
three approximations (along the row) and three temperatures
(down the columns).  
The first column shows the HFBdG based result for  
$A(\vec k,\omega)$, while the
second column shows $A^L_{trap}(\vec k,\omega)$ 
based on  $P_{MC}(n)$ 
integration.
The third column shows $A^L_{trap}(\vec k,\omega)$ 
based on $P_{LDA}(n)$ on a 
$24 \times 24$ lattice, the fourth column shows the
result on 
a $192 \times 192$ lattice. For
the larger lattice the corner potential is kept at $V_c=3t$,
as in the small system, so that the larger and smaller systems are
roughly equivalent.
All the main 
features of the HFBdG based calculation in column one survive in
the $P(n)$ based result, provided an accurate reference is used
for the flat system.

\begin{center}
\textbf{V.~Size dependence of momentum resolved spectral function.}
\end{center}
In this section we present the size dependence of $A(\vec k,\omega)$.
We demonstrate that the system size chosen in the main text is sufficient
to capture the physics.
\begin{figure*}[h]
\centering
\includegraphics[width=15cm,height=18cm]{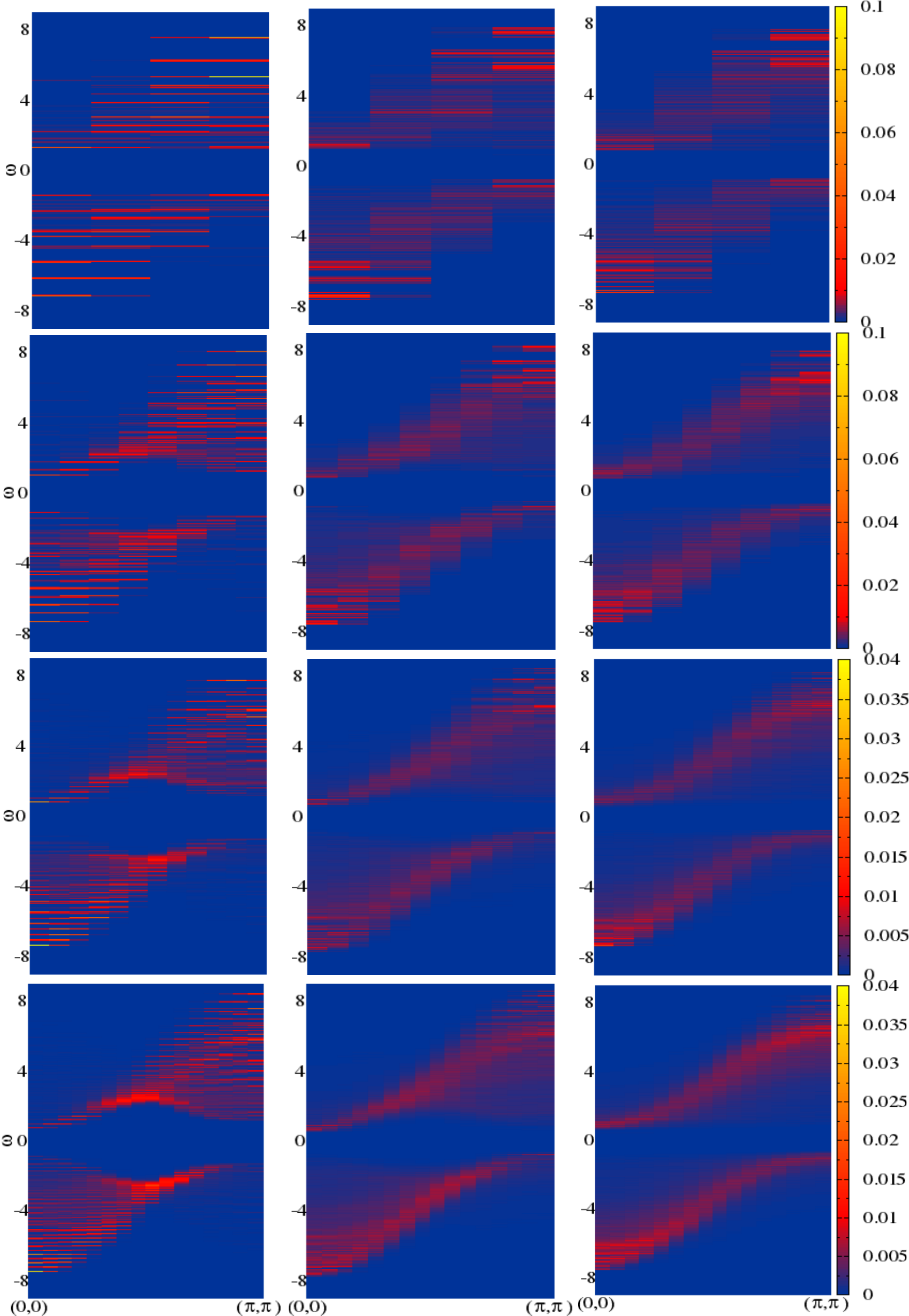}
\caption{Comparison of spectral function $A({\bf k}, \omega)$ at
different  system sizes and temperatures. 
The rows, from top to bottom, are for
sizes $8 \times 8$,  $16 \times 16$,
$24 \times 24$, and $32 \times 32.$ 
The temperature from left to 
right is $T/t=0,~0.08,~=0.3$.  
In all cases trap potential at the corner is set at
$V_c/t =3.0$ to ensure proper trap size scaling.
Other electronic parameters remain the same as in the 
paper.
} 
\label{finite-size-A}
\end{figure*}
We have computed (see Fig \ref{finite-size-A} of this reply) the spectral
function on four sizes, $8 \times 8$,
$16 \times 16$, the original $24 \times 24$, and $32 \times 32$.
We can go to even larger sizes, but these will already make the 
point. To compare results at different system sizes in this
inhomogeneous system we have kept the `corner potential' fixed
at $V_c = 3t$.  The rows, from top to bottom,
 are for $L=8,~16,~24,~32$.
The columns, from left to right, are for 
temperatures, $T/t=0,~0.08,~0.30$ as in the paper.

\vspace{.2cm}

The intent is to identify an unusual `forward bending' feature 
in $A({\bf k}, \omega)$, 
in the confined system, as the momentum crosses
$k_F$.

\vspace{.2cm}

Looking at the first column, $T=0$,
it is impossible to infer anything reliable about backbending
or `forward bending' from the $L=8$ result. 
At $L=16$ the hints are already clear. At $L=24$ the features
observed at $L=16$ are more prominent, and the wider gap at
${\bf k} \sim \{\pi/2, \pi/2\}$, compared to
${\bf k} = \{0,0\}$, is convincing. $L=32$ is 
indistinguishable from $L=24$.

\vspace{.1cm}
The persistence of this feature to intermediate temperatures is
again clear in $L=24$ and $L=32$, somewhat ambiguous in $L=16$,
and non existent in $L=8$. 

\vspace{.2cm}

So, (i)~the system size matters: a small size calculation 
would not observe the effect, and (ii)~our sizes
are large enough for the result to be scaled up to experimental
lattices (with proper trap scaling).


\begin{thebibliography}{99}
\bibitem{footnote1}
For multichannel decomposition the sum of squares of couplings in the
two channels should equal unity.
\end{thebibliography}
\end{document}